\documentstyle[aps,psfig,epsfig]{revtex}
\def\be{\begin{equation}}
\def\bea{\begin{eqnarray}}
\def\ee{\end{equation}}
\def\eea{\end{eqnarray}}

\def\ov{\overline}
\def\ra{\rangle}
\def\la{\langle}
\def\r{\right}
\def\l{\left}

\def\a{\alpha}

\def\d{\delta}
\def\C{{\bf C}}
\def\S{{\bf S}}
\def\t{\tau}

\begin{document}

\twocolumn

\title{Structurally constrained protein evolution:
results from a lattice simulation.}
\author{
Ugo Bastolla$^{(1)}$, 
Michele Vendruscolo$^{(2)}$ and
H. Eduardo Roman$^{(3)}$}
\address{$^{(1)}$FU Berlin, Inst. f\"ur Kristallographie,
Takustr. 6, D-14195 Berlin, Germany\\
$^{(2)}$Department of Physics of Complex Systems, Weizmann Institute of Science, 
Rehovot 76100, Israel \\
$^{(3)}$INFN, Sezione di Milano, Universit\`a di Milano, Via Celoria 16, 
I-20133 Milano, Italy} 

\address{
\centering{
\medskip\em
{}~\\
\begin{minipage}{14cm}
{}~~~
We simulate the evolution of a protein-like sequence 
subject to point mutations, 
imposing conservation of the ground state, thermodynamic
stability and fast folding. 
Our model is aimed at describing neutral evolution of natural proteins.  
We use a cubic lattice model of the protein structure
and test the neutrality conditions by extensive Monte Carlo simulations.
We observe that sequence space is
traversed by neutral networks, {\it i.e.} sets of sequences with the
same fold connected by point mutations. Typical pairs of sequences on a neutral
network are nearly as different as randomly chosen sequences. The
fraction of neutral neighbors has strong sequence to sequence variations,
which influence the rate of neutral evolution.
In this paper we study the thermodynamic stability of different protein
sequences.
We relate the high variability of the fraction of neutral mutations
to the complex energy landscape within a neutral network, 
arguing that valleys in this landscape
are associated to high values of the neutral mutation rate.
We find that when a point mutation
produces a sequence with a new ground state,
this is likely to have a low stability. 
Thus we tentatively conjecture that neutral networks 
of different structures are typically well separated in sequence space.
This results indicates that changing significantly
a protein structure through a biologically acceptable chain of point
mutations is a rare, although possible, event.
{}~\\
{}~\\
\end{minipage}
}}

\maketitle

\section{Introduction}
\label{sec:introduction}
Almost unmistakingly, naturally occurring proteins with sequence similarity
larger than 40\% adopt similar folds \cite{rost99}.
Since natural proteins with homologous sequences descend from a common
ancestor, this observation indicates that protein structures
are significantly conserved in evolution. 
Indeed, several proteins with different functions show a remarkable
structural similarity of evolutionary origin even if their sequences
can not anymore be recognized as related \cite{holm,qasba}.
A recent study on the Protein Data Bank (PDB) 
showed that the typical sequence similarity
between proteins with the same fold is about 8.5\% \cite{rost},
only slightly larger than for a random pair of sequences \cite{random}.
In this set also proteins with common ancestors are likely to exist.
These observations cue to the fact that during evolution, 
there is a strong memory for the structure 
but only a very loose memory for the sequence.

The neutral theory of molecular evolution, proposed 
in 1968 by Kimura \cite{Kimura} and, independently, by Jukes and King
\cite{JK}, is well consistent with these observations.
Kimura suggested that most amino acid substitutions in protein sequences
are selectively neutral, {\it i.e.} indistinguishable from the
wild type from the phenotypic point of view, and are 
fixed  by chance in biological populations \cite{Kimura}. This hypothesis
has been heatedly debated in the genetic literature \cite{Gillespie,Ayala}.
Strictly speaking, conservation of the fold and neutral evolution
are not equivalent, since neutrality deals with the activity of the protein,
concentrated in its active site, more than with its structure. Moreover
drastic changes in the environment can modify the selective value of protein
structures. 
However, since our model does not represent biological activity, we
assume in the following
neutrality to be synonymous of structure conservation.

In the last decade, the possible occurrence of neutral evolution
has been revealed by a series of computational and analytic studies
of the sequence to secondary structure relationship for RNA molecules
\cite{RNA}. An exponentially large number of sequences correspond
on average to a single structure, and the distribution of the number of
sequences per structure is quite broad (following a power law), with the most common
structures forming connected neutral networks which percolate sequence space.

For proteins, the sequence to structure relationship is much more difficult
to study than for RNA. Shakhnovich and Gutin \cite{SGneut}, using the
Random Heteropolymer model, argued that the probability that a point mutation
is neutral ({\it i.e.} it does not alter the native state) is non vanishing
even for very long sequences. 
In the same spirit, Tiana {\it et al.} \cite{Tiana}
considered a cubic lattice model and a sequence with 36 residues, optimized
in such a way that its ground state coincides with a target structure and
is very stable \cite{S36}, and estimated that 70\% of the point mutations
are neutral. Bornberg-Bauer \cite{BB} studied a two-dimensional HP model
with only two residue types and chain length $N=18$
by using exact enumeration. He found, in analogy
with the RNA case, that the distribution of the number of sequences per
structure is very broad, but sequences corresponding to the same structure
are clustered in small regions of sequence space.

We simulate the evolution of a protein sequence subject to structure
conservation. Mutations that change the protein's native structure, identified
with the ground state of the model, are considered lethal and are rejected.
In this way,
our sequence follows an evolutionary trajectory on a neutral network,
{\it i.e.} a set of sequences sharing the same fold and connected by point
mutations. 
While the structure (fold) is conserved, the sequence changes 
as new mutations are accepted, and after a sufficient number of steps 
along the evolutionary trajectory have been performed, the sequence behaves
essentially as a random one with respect to the original one \cite{UME}.
 
It is important, however, to impose not
only the condition that the native state is conserved but also that
its stability remains high and the folding time remains low. These
conditions are not only biologically relevant, 
but also help the model protein to diffuse in sequence space.

We use in this study a lattice representation of protein conformations,
because only in this way the ground state and its thermodynamic stability
can be reliably determined, but we believe that our
simplified model reproduces the generic features of the evolution of real
protein sequences.

Support to our results comes from a recent 
study by Babajide and coworkers \cite{stadler}, who
found evidence for the presence of neutral networks in protein sequence space.
Their work is similar in spirit to the present one, 
but rather different methodologically.
Real protein structures were represented through the $C_\alpha$ and 
$C_\beta$ coordinates taken from the PDB, and an approximate criterion of
fold recognition based on the Z score \cite{Z} was used.
Further support also comes from the work of Govindarajan and Goldstein
\cite{gold1,gold2}, who introduced the ``foldability'' landscape
in order to describe molecular evolution. 
In the language of Govindarajan and Goldstein
the foldability of a protein represents its fitness 
for survival during evolution and it is related to the stability 
and to the kinetic accessibility of the native state. 
Govindarajan and Goldstein also found that their evolutionary
dynamics in sequence space was confined inside ``neutral networks''.

Tha main result of our paper, namely the fact that the fraction of
neutral neighbors strongly fluctuates inside the neutral network, and that
these fluctuations can be related to the foldability landscape,
should also be put in relation with the recent preprint by Tiana {\it et al.}
\cite{Tiana2} where it is shown that the energy of a target structure
has a complex landscape with valleys and barriers in sequence space. The
present work supports such a picture by using the same protein model but 
employing rather different methods of investigation.

We already presented some results on neutral networks in Ref. \cite{UME}.
Here we focus our attention on the issue of the stability of the native
state, relating it to the characteristics of the evolution.
In Sec.\ref{sec:model}, we describe our model
protein and our protocol to simulate neutral evolution. 
In Sec.\ref{sec:NN} we summarize our previous results.
In Sec.\ref{sec:sequences} 
we describe the properties of the sequences generated, dividing
them in four classes. This section focuses on the relation between
thermodynamic stability and evolutionary dynamics.
Sec.\ref{sec:discussion}  presents an overall discussion, relating our results
to biological observations.

\section{A simple model of protein evolution}
\label{sec:model}

In this section we define the lattice model
used to represent protein structure and the algorithm introduced
to simulate evolution in sequence space.

\subsection{Lattice model of protein structure}
To investigate the correspondence between sequences and structures
we use a lattice model with twenty amino acid types.
We consider sequences of length $N=36$,
denoted by the symbol
$\S =\{s_1,\ldots,s_N\}$, where $s_i$ belongs to a twenty-letter alphabet. 
Configurations are represented
by self avoiding walks on the simple cubic lattice, where each occupied
site represents an amino acid.
An energy $E(\S,\cal{C})$ is assigned to configuration $\cal{C}$ of
sequence $\S$ according to the rule:

\be 
E(\S,\C)=\sum_{i<j}^{1,N}C_{ij} U(s_i,s_j) \;,
\ee
where $U(a,b)$ is a $20\times 20$ symmetric interaction matrix expressing
the contact interactions of amino acids of species $a$ and $b$.
We use an interaction matrix $U(a,b)$ derived from the Miyazawa-Jernigan
interaction matrix \cite{MJ}.
The matrix $\C=\{C_{ij}\}=f(\cal{C})$, called the {\it contact map} of
configuration $\cal{C}$, has elements $C_{ij}$ equal to one if
residues $i$ and $j$ are nearest neighbors
on the lattice but not along the chain and zero otherwise.
The similarity between contact maps is measured through the overlap
$q(\C,\C^\prime)$, defined as

\be 
q(\C,\C^\prime)={1\over N_c^*}\sum_{i<j}C_{ij} C^\prime_{ij}\;,
\ee
where $N_c^*$ is the maximal between $N_c$ and $N_c'$, the
number of contacts respectively of two contact maps $\C$ and $\C^\prime$,
and $N_c=\sum_{j>i}C_{ij}$. With this definition, two maps are identical if
and only if $q=1$. Note that this does not imply in general identity of
configurations. Nevertheless, we use the
overlap as a measure of similarity in configuration space because
structures with the same contact map are degenerate in energy 
and for compact structures, only small conformational fluctuations 
are allowed when the entire set of contacts is specified.
Moreover, such structural fluctuations might play an important 
role in protein functionality \cite{wright}.

The native structure of sequence $\S$ is identified with the
ground state of the model if it is thermodynamically stable.
We evaluate stability by measuring the average overlap $\la q\ra$ between
the ground state and the Boltzmann ensemble of structures:

\be
\la q\ra={1\over Z}\sum_{\cal C} q(\C_0,f({\cal C})) e^{-E({\cal C},\S)/T},
\ee
where $\C_0$ is the contact map of the ground state, $f(\cal{C})$ is the
contact map of configuration $\cal{C}$ and $Z$ is the partition function.
This quantity is close to one if all the low energy structures are quite
similar to the native state. In this case the energy landscape of the
model is well correlated, and the sequence is also expected to be a good
folder.

\begin{figure}
\caption{The ``native state'' of the model protein.
The initial residue is the one in the bottom left corner.}
\label{fig:conf}
\end{figure}

We consider the target contact map $\C^*$ represented in Fig.\ref{fig:conf}.
It has $N_c=40$ contacts, the maximal number of contacts
possible for a chain of length $N=36$. In this case the contact map defines
uniquely the configuration of the system.
The contact map $\C^*$ was studied by Shakhnovich and coworkers 
in a computer experiment of inverse folding \cite{S36}. 
They designed a sequence
with ground state on $\C^*$ using the procedure of Ref. \cite{SGinv},
and showed that $\S^*$ has good properties of kinetic foldability and
thermodynamic stability at the temperature where the folding is fastest.
The lower part of the energy landscape of this
sequence is remarkably smooth:
all the structures with low energy have a high overlap $q$ 
with the ground state. The lowest energy of configurations with a
fixed value of $q$ decreases regularly as $q$ approaches one. 
This correlated energy landscape,
reminiscent of the ``funnel'' paradigm \cite{BW}, 
is the reason of the good folding properties of the sequence, 
which is very different from a random sequence. 
In \cite{Tiana} it was shown that the same sequence is also very
stable against mutations. It was estimated that
about 70\% of the point mutations performed on $\S^*$
result in new sequences with exactly the same ground 
state and good folding properties. 
Thus energy minimization makes $\C^*$ stable not only in structure
space, but also in sequence space.

We note that $\C^*$ is an atypical structure 
for the interaction parameters that we choose: 
since $U(a,b)$ has average value zero and variance $0.3$, one would
expect open structures to be energetically favored. 
Indeed, typical random sequences with $N=36$ and Gaussian contact interactions
whose average vanishes have a ground state with approximately 29-33 
contacts \cite{ugo}, being thus less than maximally compact. 

\subsection{Sequence space}
In this study we consider only point mutations, thus all sequences
have the same length $N=36$ and the metric in sequence space is
given by the Hamming distance,
\be
D(\S,\S') = \sum_{i=1}^N \left[ 1 - \delta (s_i,s'_i) \right ] \; ,
\ee
where $\d$ is the Kronecker symbol and $s_i$ takes 20 different values,
one for each amino acid. A measure of sequence similarity is then given
by the overlap $Q(\S,\S')$,
\be 
Q(\S,\S')={1\over N}\sum_{i=1}^N \d(s_i,s'_i) \;,
\label{eq:dh}
\ee
which is equal to one minus the normalized Hamming distance.

We introduce also the distance $D_{HP}(\S,\S')$ and the overlap
$Q_{HP}(\S,\S')$ to measure differences in hydrophobicity.
These are defined by transforming every sequence into
a sequence of binary symbols, either H or P, 
according to the hydrophobicity of the residue.
We consider 8 hydrophobic amino-acids and 12 polar ones. 
The definitions of $D_{HP}$ and $Q_{HP}$
are analogous to those of $D$ and $Q$, 
where now $s_i$ can take only two values.

\subsection{Evolutionary process}

Our protein sequence evolves through point mutations subject to
conservation of the target contact map $\C^*$, representing the biologically 
active native structure \cite{ags96}. We impose this condition by simulating
the following iterative procedure:

\begin{enumerate}
\item
At $t=0$ we start from $\S(0)=\S^*$, which has $\C^*$ as its
``native state''.
\item
At time $t$ we mutate at random one amino acid in $\S(t-1)$,
producing a new sequence $\S'(t)$. 
\item
We submit the new sequence to selection according to the criteria
specified below.
If the sequence is accepted then $\S(t)=\S'(t)$,
otherwise we restore $\S(t)=\S(t-1)$.
\end{enumerate}
The selection step is governed by the following three conditions

\begin{description}
\item[Conservation of the ground state].
The ground state $\C$ of $\S'(t)$ must have an overlap with $\C^*$
equal or larger than a given ``phenotypic threshold'' $q_{\rm thr}$.
\be 
q(\C^*,\C) >  q_{\rm thr} \;, 
\ee
In our calculations we imposed strict conservation of the native state
by setting $q_{\rm thr}=1$.

\item[Thermodynamic stability]. 
We define thermodynamic stability through the condition
\be 
\l\la q(\C^*,\C)\r\ra > \la q\ra_{\rm thr} \;, 
\ee
where $\la \cdot\ra$ represent a Boltzmann average at the temperature $T$ 
of the simulation and $\la q\ra_{\rm thr}$ is a fixed parameter.
This condition implies that all the thermodynamically relevant states
are very similar to the target state.

\item[Kinetic accessibility].
The structure $\C^*$ must be reached in a limited number of steps of our
Monte Carlo algorithm, in at least two independent attempts.

\end{description}
For the test of the sequences
we used the PERM method \cite{PERM,Helge}, a Monte Carlo algorithm
particularly suited for finding the ground state of lattice polymers.  
Note that there is no bias towards $\C^*$ in our Monte Carlo algorithm,
{\it i.e.} it has the same a priori probability of being visited as any other
possible structure.
We remark that other schemes of simulations are also suitable 
to the same effect, as e.g. the Monte Carlo algorithm used 
in Ref. \cite{Tiana3}.
Such Monte Carlo method with moves in configuration space is more
suitable than PERM to estimate folding times. However, due to
computational limitations, we did not try to measure accurately the folding
time, thus we adopted the PERM method, which is faster for the task that is
interesting for us.

The test of a new sequence $\S$ is divided into three phases:

\begin{itemize}
\item
We discard $\S$ if after $m$ iterations $\C^*$ is not reached
or if other structures of energy lower than $\C^*$ are found.
\item
Otherwise we continue to run the algorithm for another $m$ iterations
and discard $\S$ if we find structures of energy lower than $\C^*$.
\item
If $\S$ passed the first two phases, we run again and independently
the MC algorithm for a time $2m$ and accept $\S$ if also this time $\C^*$
is found as the lowest energy structure.
\end{itemize}

Thus for each accepted sequence we run the algorithm for $4m$ steps,
with $m=124000$.
We never found in the second independent run of the MC
algorithm a structure with lower energy than the putative
ground-state $\C^*$ found in the first run.
This fact encourages us to believe that the
algorithm was effective in finding the ground state. Another
support to this conclusion comes from the fact that, as it will be
discussed later in more detail, all of the
selected sequences have a remarkably correlated energy landscape, which
makes the task of finding the ground state easier.

On the other hand, whenever the sequence was rejected, 
we are less sure that we were able to determine its ground
state. The difference is due to two reasons: first, we investigate rejected
sequences on the average for a shorter time. Second, rejected
sequences have typically a less correlated energy landscape, so
that the determination of the ground state should be more
difficult. Nevertheless, we shall present in Sec.\ref{sec:sequences} also
data about rejected sequences, since they are interesting and
refer to a very large number of sequences, even if they are
individually not completely reliable. 

\vspace{.5cm}
The three conditions for the acceptance of a mutation enforce
the conservation of the fold of the protein. This is similar to neutral
evolution where the biological activity of the mutated sequence does not vary.
Nevertheless, conservation of the fold is not a necessary condition
for selective neutrality in real proteins, although
a very high degree of conservation is usually observed,
and it is not even a sufficient one, 
since - in the case of enzymes - the active site has also to be
conserved and the environment has to remain reasonably stable.
Thus our model represents the neutral evolution of the part of the chain not
involved in chemical activity, in a stable chemical
environment. Despite its simplifications, we believe that our model captures
important features of structural constraints in the neutral evolution
of proteins.

\section{Neutral networks}
\label{sec:NN}

In this section we summarize results regarding the diffusion
in sequence space under our model of neutral evolution. More details have
been given in Ref. \cite{UME}.

\subsection{Hamming distance}
\label{sec:distance}

An interesting result of our simulation is that sequences originated
from the same common ancestor diverge so much that their similarity
is almost as low as for a random pair of sequences while their
structures remain unchanged.
Starting from the same sequence we generated eight realizations of
neutral evolution, simulating the phylogenetic radiation of eight species
from a common ancestor. We use the following values of the
selection parameters: phenotypic threshold
$q_{\rm thr}=1$, corresponding to exact conservation of the ground state,
stability threshold $\la q\ra_{\rm thr}=0.90$ at a temperature $T=0.16$
chosen so that the folding of the initial sequence $\S^*$ is fastest
with our Monte Carlo algorithm.

\begin{figure}
 \centerline{\psfig{file=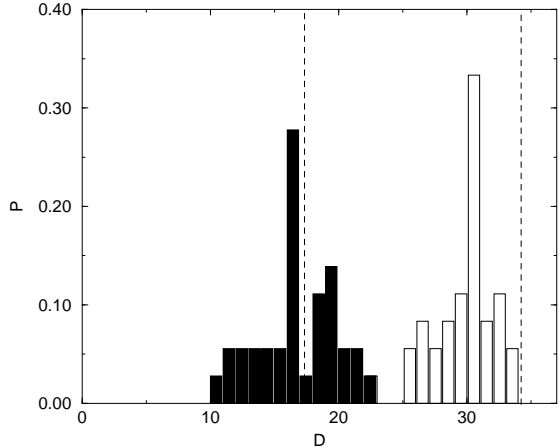,height=7cm,angle=0}}
\caption{Histograms of the Hamming distances between the end points of
  eight independent trajectories using the full 20 amino-acid alphabet
  (white) and the reduced HP alphabet (black). The vertical dashed lines
  represent the average values for random pairs of sequences.}
\label{fig:endpoints}
\end{figure}

The average Hamming distance between the final points of the eight
evolutionary trajectories is $D=30.2$, only slightly smaller than the
random value $D^{ran}=34.2$ (see Fig.\ref{fig:endpoints}).
However, this quantity had not yet reached
a stationary value when the simulations were interrupted, thus
we can not exclude that the long time behavior coincides with $D^{ran}$.
An indication in this sense is the fact that the maximum distance
between sequences in two different trajectories is $D=35$.
All the residues in the original sequence could be substituted
at least twice, but some are more difficult to change.
We define the degree of conservation, or rigidity, of residues at the
$i$-th position in the sequence as follows:
\be 
R_i=\sum_a P_i^2(a)  \;,
\ee
where $P_i(a)$ is the probability to find the amino-acid $a$ at position $i$.
$P_i(a)$ is estimated from the end points of the eight neutral paths generated.
$R_i=1$ if the amino-acid at position $i$ is never changed, 
while $R_i\approx 1/8$ if it is completely random. We found that several
positions have rigidity compatible with the random value, and no position
has $R_i=1$. As one would expect, the most conserved positions are the two
in the interior of the structure (see Fig.\ref{fig:endpoints}).

The same results hold for the HP representation in which hydrophobic (H)
and polar (P) amino acids are grouped together so that $s_i$ can assume
only two values.
The average value of the distance is in this case
$D_{HP}=16.3$, not far from the random value $D_{HP}^{ran}=17.3$, and 
the variance is $V_{HP}=6.0$, compatible with $V_{HP}=D_{HP}(1-D_{HP})/N$.
It is at first sight surprising that also 
the distance $D_{HP}$ is close to that expected for random
sequences. This is in part an effect of the short length of our sequences,
since only two residues are in the interior of the structure, while all other
ones are at the surface. 
It is interesting to note, however, that also the two
residues in the interior of $\C^*$ have rigidity $R_i<1$,
even when the two letter HP representation is used. 
The distinction between polar and hydrophobic
residues is based only on the interaction matrix that we use, in which also
polar residues can have attractive interactions.
It is interesting to note that a recent study of real protein
structures \cite{ms99} found a correlation between amino acids buried in the
core and evolutionary conserved ones, consistently with our results.

\subsection{Neutral Mutation Rate}

For a given sequence $\S$ of $N$ amino acids,
we define the neutral mutation rate $x(\S)$ as the fraction 
of acceptable non-synonymous mutations
\be
x(\S)= \frac{1}{20N}\sum_{i=1}^{N}\sum_{\alpha\neq s_i}^{1,20}
\chi_{\alpha i}(\S)\; , \label{eq:x}
\ee
where $\chi_{\alpha i}(\S)$ equals one if assigning the amino acid of species
$\alpha$ at position $i$ on the sequence $\S$ does not change the native state,
and zero otherwise. 
Non-synonymous mutations are those
for which an amino acid is not replaced by itself.

The simplest measure of the neutral mutation rate
is obtained by computing the frequency of neutral mutations over all the
non-synonymous mutations proposed. In this way we found
$\ov{x}\approx 0.05$ (the overline represents an average over
the mutational process). However, this quantity alone is not enough to
characterize $x(\S)$, which fluctuates strongly in sequence space. For
instance, it was estimated by one of us and coworkers \cite{Tiana} that
$x(\S^*)\approx 0.7$, where $\S^*$ is the starting point of our
evolutionary trajectories.

We measured indirectly the distribution of $x(\S)$ in sequence space
from the distribution of the ``trapping'' time $\t_t(\S)$ 
that a trajectory spends on sequence $\S$. The average value of the
trapping time is inversely proportional to the neutral mutation rate: 
\be
\ov{\t_t(\S)}=\frac{1}{x(\S)} \; ,
\ee
where the bar denotes average over the different mutations. The
distribution of $\t$ at fixed $x$ is a geometric one,
$P_x(\t)=x(1-x)^{\t-1}$,
so that, averaging over the neutral set, we get
\be 
\l[P(\t)\r]=\int_0^1 dx~p(x)\l(x\over 1-x\r)(1-x)^\t \; \label{eq:Ptau},
\ee
where $\l[\cdot \r]$ denotes an average over sequences belonging to 
the neutral network (in this argument we neglect
the error in evaluating whether a sequence
belongs to the neutral set: in particular, the conditions of fast
folding and of thermodynamic stability are subject to considerable
evaluation errors). 

\begin{figure}
  \centerline{\psfig{file=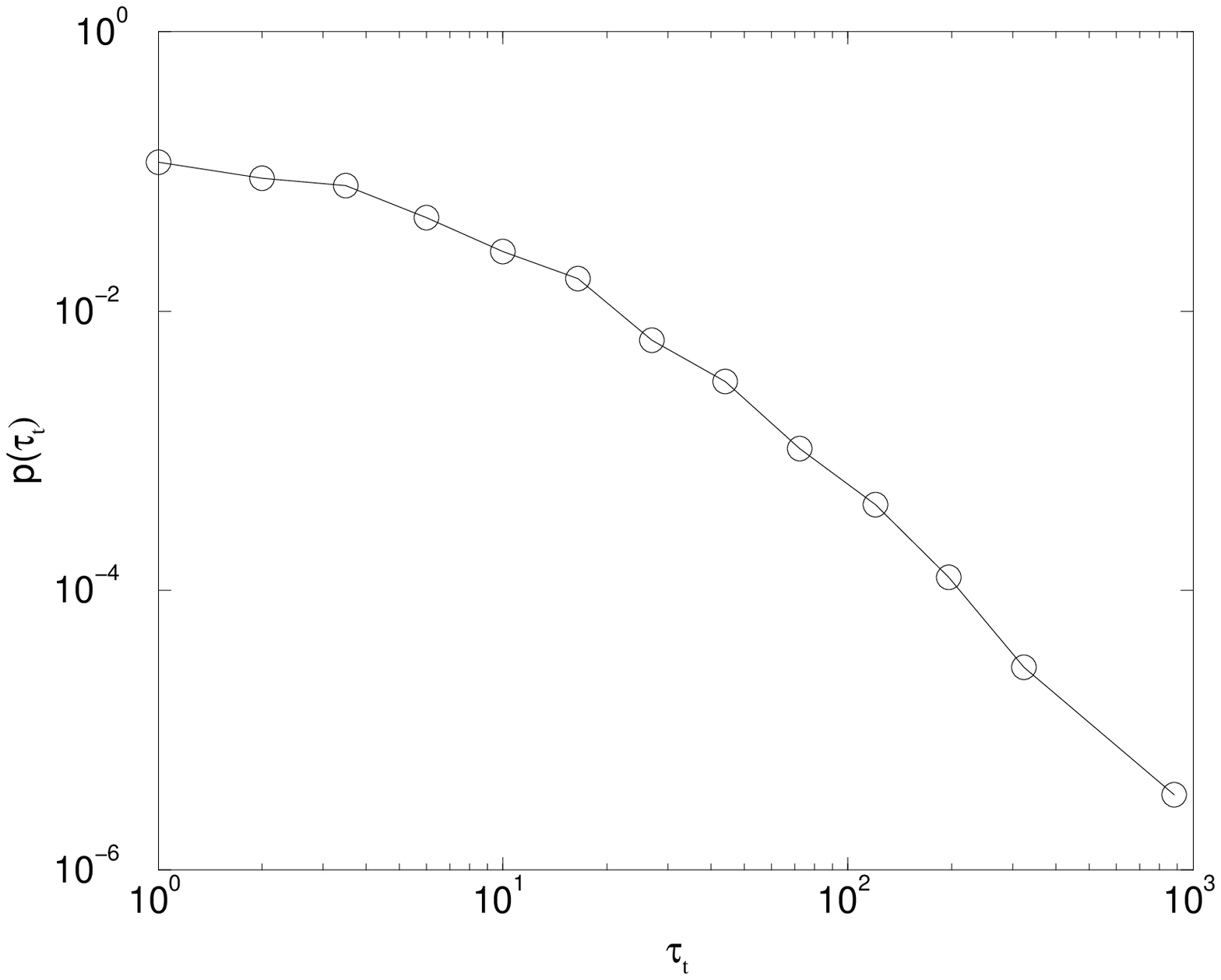,width=7cm,angle=0}}
\caption{Distribution of the trapping time $\tau_t$.}
\label{fig:trapping}
\end{figure}

The distribution of $\t_t$ is  broader than an exponential one
(Fig.\ref{fig:trapping}), thus, even if we can not invert Eq. \ref{eq:Ptau},
we expect that the distribution of the neutral mutation rate $x$ is
also broader than exponential.

The values of $\t_t$ for neighboring sequences are rather correlated,
but the correlation seems to vanish after few steps in sequence space
(data not shown).

\subsection{Genetic drift and population genetics}

In Kimura's neutral theory \cite{Kimura} it is assumed that the fraction of
neutral mutations $x(\S)$ does not depend on the sequence. With this
hypothesis, the time evolution of the Hamming distance $D(t)$ between the
starting sequence $\S(0)$ and the present sequence $\S(t)$ is given in our
model by
\be 
D(t)/N\approx \l(1-{1\over 20}\r) \l[1-\exp(-xt/N)\r]\;,
\label{eq:xexp}
\ee
where the time $t$ represents the number of mutational events.
However, this hypothesis is contradicted by our results, which show
that the relaxation of the distance is not exponential. This fact is due to
the large fluctuations of the neutral mutation rate along the neutral network.

This qualitative result can be interesting for the understanding of
protein sequence evolution.
A key issue in the theory of molecular evolution is the following:
Is the rate at which amino acids are substituted in protein sequence
constant on different branches of the phylogenetic tree?
The constancy of the substitution rates was proposed by
Zuckerkandl and Pauling in their pioneering study of molecular evolution
as the molecular clock hypothesis \cite{ZP}. This hypothesis has been
questioned recently \cite{Gillespie,Ayala}, even if it seems to be at least
approximately valid for several proteins.

Kimura's theory predicts that neutral substitutions occur at a rate
$r=\mu x$ which is the product of the bare mutation rate $\mu$ times the
fraction $x$ of neutral mutations. This rate is independent of the
size of the population. The value $x$ characterizes the substitution rate of
a particular protein but does not vary during evolution.
Mutations occuring in a geological time $T$ are thought to follow a Poissonian
statistics with average value $\mu T$, so that the number of
substitutions is predicted to have Poissonian statistic with
average value $\la n_S\ra=\mu Tx$. This has the
important consequence that the fluctuations of the substitution process
in different species should increase only as $\sqrt{T}$. More precisely,
the ratio $R(T)$ between the variance and the average value of the number of
substitutions should be identically equal to one. This strong prediction
was first tested by Kimura \cite{Kimura} with the conclusion that deviations
from the Poissonian statistics are small. However, more recently
Gillespie repeated the test for a larger number of proteins, finding that
for most of them the value of $R(T)$ is much larger than one. He thus
argued that the hypothesis that most mutations are neutral has to be
rejected.

Our results provide an alternative
explanation: the strong fluctuations of the substitution process can be
attributed to the fluctuations of the neutral mutation rate in sequence
space, even in the absence of any selective pressure. 
To test this hypothesis, we assume, as above, that the number $m_k(T)$ of
attempted mutation events in a time $T$ during trajectory $k$
is a Poissonian variable of average value $\mu T$.
In the present study, $k=1,...8$ is the label of the evolutionary trajectories.
Then for every trajectory $k$ we count the number $n_k(T)$
of mutations accepted over $m_k(T)$ steps of our evolutionary algorithm. 
This number is then interpreted as the number 
of substitutions in ``species'' $k$.
We can thus compute the variance and the average value of this variable
over the eight trajectories. The ratio between them gives an estimate of
the dispersion ratio $R(T)$. This is always larger than one, contradicting
the Poissonian hypothesis. Moreover, $R(T)$ 
is found to be an increasing function of $T$, 
so that it is no longer true that the fluctuations grow with time as $\sqrt T$.

Since our model takes into
account only neutral and lethal mutations, without considering either
advantageous mutations or slightly deleterious ones, we conclude that
the violation of Poissonian statistics is not a decisive proof against the
validity of the neutral hypothesis.

\section{Properties of the sequences}
\label{sec:sequences}

We classify the nearly 12,000 sequences generated by our evolutionary algorithm
in four classes, with the reminder that the identification of the ground
state is only tentative for rejected sequences, as already discussed.

\begin{enumerate}
\item {\bf Selected sequences}, belonging to the neutral network.
Their number is a fraction $f=0.050$ of the total set.

\item {\bf Unstable sequences}, $f=0.172$.
Their lowest energy state coincides
with $\C^*$, but the stability condition is not fulfilled. The
rejection was made in most cases already after the first MC run, if the
condition $\l\la q(\C,\C^*) \r\ra > 0.75$
was not fulfilled, otherwise the sequence was studied in another MC run.

\item {\bf Slow folding sequences}, $f=0.472$.
For such sequences, no structure
with energy lower than $E(\C^*,\S)$ was found, but the MC algorithm
did not reach the target structure $\C^*$. In some cases
($f=0.061$) $\C^*$
was reached in the first MC run but not in the second one, while in
most cases the rejection was made already after the first run.

\item {\bf Structurally mutated sequences}, $f=0.306$.
For such sequences the
lowest energy structure $\C_0$, has lower energy than the target structure:

\be E(\C_0,\S)<E(\C^*,\S). \ee

\end{enumerate}

Before describing separately the properties of these classes of sequences,
we show that the value of the $Z$ score is able to
distinguish statistically the different classes.
The $Z$ score \cite{Z} is used to evaluate the match between a sequence $\S$
and a structure $\C^*$ taken from a pool of alternative structures. It is
defined as

\be Z(\C^*,\S)={E(\C^*,\S)-\l\la E(\C,\S)\r\ra \over
\sqrt{\l\la E^2(\C,\S)\r\ra-\l\la E(\C,\S)\r\ra^2}}, \ee
where the brackets denote average with respect to the ensemble of alternative
structures at high temperature. The more negative $Z$ is, the better is the
match between the sequence $\S$ and the structure $\C^*$ and the more stable
is the structure $\C^*$, provided that it is really the lowest energy
structure. This measure is often used in
computer experiments of fold recognition \cite{stadler,Z}.

Following Mirny and Shakhnovich \cite{MS}, we use a simplified measure
of the $Z$ score, considering as alternative
structures only maximally compact structures and approximating their
average energy and their variance with, respectively, the average energy 
and the variance of the set of all possible contacts (we take into
account the fact that in the simple cubic lattice the only possible contacts
are those between monomers of different parity). More precisely, our definition is

\be Z'(\C^*,\S)={
E(\C^*,\S)-Nc_{max}\l[ U(S_i,S_j)\r]
\over
Nc_{max}\sqrt{\l[ U^2(S_i,S_j)\r]-\l[ U(S_i,S_j)\r]^2}
}, \ee
where
\be U(S_i,S_j)={\sum_{ij}P_{ij}U(S_i,S_j)\over
\sum_{ij}P_{ij}}, \ee
$P_{ij}$ is one if a contact between amino acids $i$ and $j$ is
possible in some configurations, zero otherwise and $Nc_{max}$ is the
number of contacts for maximally compact structures, $Nc_{max}=40$ for
$N=36$ on the cubic lattice.
$Z'$ is a good approximation to the $Z$ score and it is very easy
to compute numerically, without the need for a simulation at a high
temperature.

\begin{figure}
\centerline{\psfig{figure=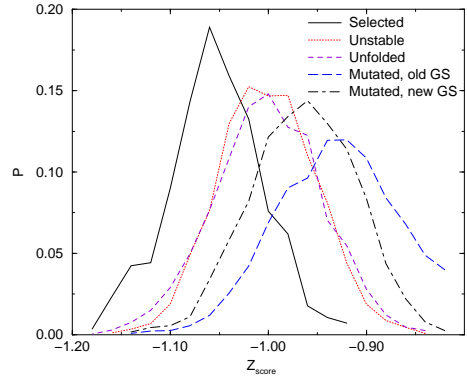,width=7.0cm,angle=0}}
\caption{Distribution of the $Z$ score for different classes of sequences.}
\label{fig:zscore}
\end{figure}

We plot in figure \ref{fig:zscore} the distribution of the $Z$ score for the
four classes of sequences. For the structurally mutated sequences
we evaluated both the $Z$ score of the target structure $\C^*$ and the
$Z$ score of the lowest energy structure found, $\C_0$. Note that the
starting sequence $\S(0)$ has the lowest value of the $Z$ score, i.e.
$Z=-1.22$.

The ranking of the $Z$ score for different classes is as expected on the basis
of their stability. The most negative values of $Z$ are proper of selected
sequences, which are most
stable. Next come slow folding and unstable sequences. The $Z$ score of
mutated ground states, $Z(\C_0,\S)$, are less negative. Last
in this rank of stability comes the $Z(\C^*,\S)$ for the
structurally mutated sequences. In this case, we are sure that $\C^*$
is not the ground state of the sequence.

\begin{figure}
\centerline{\psfig{figure=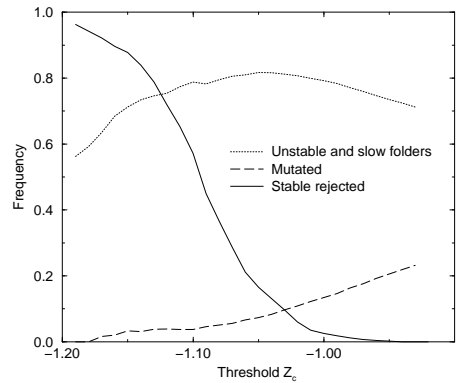,width=7.0cm,angle=0}}
\caption{Fraction of unstable, slow folding and mutated sequences
that would be selected with a criterion based on a threshold value of
the $Z$ score, $Z_c$. Solid line: fraction of sequences selected by our
algorithm that would be rejected with the same criterion.}
\label{fig:threshold}
\end{figure}

These results confirm that the evaluation of the $Z$ score is an efficient
criterion to decide whether $\C^*$ is the ground state of a sequence
$\S$. Nevertheless, the fact that the distributions of the $Z$ score
relative to different classes have large overlaps should make one worry 
that the $Z$ score is not a precise criterion. In Fig.\ref{fig:threshold}
we report the results that we would get using a threshold value of $Z$, $Z_c$,
as a criterion for fold recognition, instead of studying the sequences with
Monte Carlo simulations. The dotted line represents the fraction
of sequences that are unstable or slow folders according to our criterion and
would be accepted with a criterion based on the $Z$ score. This goes from
less than 60\% for the most stringent threshold to a plateau value of
about 80\%. We can say very little about this class of sequences.
It includes sequences that are really of lower quality than selected sequences,
sequences that are of the same quality but were not selected because of the
uncertainties of the selection procedure and sequences which do not fold to
the target state. The dashed line represents sequences whose ground state
is surely different from the target one. Their fraction increases from zero
to about 20\% as the threshold becomes less stringent. Finally, the solid
line represents sequences that would be selected with our criterion but not
with the $Z$ score criterion, as a fraction of the total number of selected
sequences. Our results indicate that a good choice for the
threshold could be $Z_c\approx -1.07$: 
14\% of the sequences selected with this criterion would
also be selected with our criterion, 80.5\% would be sequences that do not
fulfill the stability or fast folding conditions and 5.5\% would be sequences
which have certainly a different ground state, but probably similar to the
target one, since the $Z$ score of mutated sequences is correlated to the
similarity between the new ground state and the target state (see below).
About 29\% of the sequences that our criterion selects would be discarded
with the $Z$ score criterion. This number becomes much larger if the
threshold is made more stringent. Thus, the criterion based on $Z$ accepts
most sequences that we reject and rejects a large fraction of those that
we select.

\vspace{.3cm}
The distribution for the slow folding class is quite similar
to that of the unstable class. This is not surprising, since it is
well known that stability and fast folding are correlated
in lattice heteropolymer models \cite{KT,gap}. In particular, stability
as we defined it requires a correlated energy landscape, which is considered
a property of fast folding sequences.
Thus these results encourage us in believing that the conditions we
imposed and the algorithm to verify them were appropriate.

Interestingly, the distribution relative to $Z(\C_0,\S)$ lies
to the right of the other ones, indicating that the stability of the
structurally mutated ground states is rather
low. This is not unexpected. In fact, structurally mutated sequences
are only one point mutation apart from selected sequences, thus
$\C^*$ should still have a low energy and should decreases the stability
of $\C_0$. We shall comment further on this point in the conclusions.

\begin{figure}
\centerline{\psfig{figure=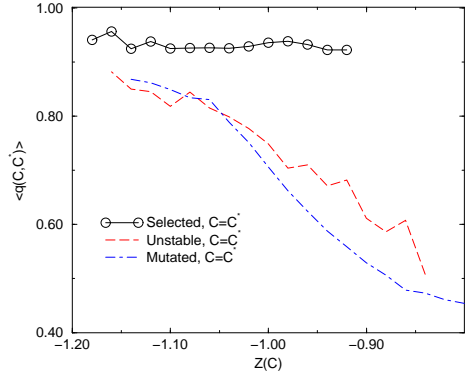,width=7.0cm,angle=0}}
\caption{Average value of the native overlap 
as a function of the $Z$ score $Z(C)$.}
\label{fig:Zq}
\end{figure}

The $Z$ score correlates well to the native overlap $\la q(\C,\C^*)\ra$ that
we assumed as a measure of thermodynamic stability (Fig.\ref{fig:Zq})
for unstable sequences (correlation coefficient $r=-0.48$) and for
mutated sequences (this is due to the fact that the overlap between the
mutated ground state and the native state correlates with to the $Z$ score).
No correlation is visible for selected sequences, which always have
$\la q(\C,\C^*)\ra>0.9$. For unfolded sequences the measure of
$\la q(\C,\C^*)\ra$ is not possible (see Fig.\ref{fig:Zq}).

\subsection{Selected sequences}
\label{sec:energy}
With our criterion we accepted 
566 sequences in six evolutive trajectories.
Such sequences are both fast folding and thermodynamically stable.
These properties are not typical of random sequences \cite{SGinv}.

\begin{figure}
\centerline{\psfig{figure=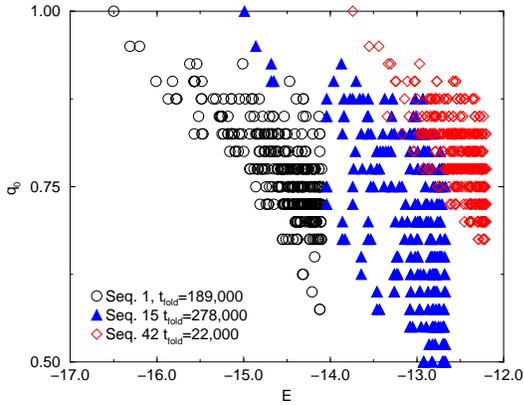,width=7.0cm,angle=0}}
\caption{Correlations between the energy and the similarity with
  the ground state for the 200 lowest energy structures of three
  selected sequences.}
\label{fig:spec}
\end{figure}

For selected sequences there is a significant
correlation between the energy $E$
of a conformation and its overlap $q$ with the native state.
In Fig. \ref{fig:spec} we represent the 500 lowest energy configurations
of three different sequences as points in the $(E,q)$ plane.
For every sequence, all points fulfill the inequality

\be 
1-E(\C,\S)/E(\C^*,\S) \geq \a(\S)\l(1-q(\C,\C^*)\r). \;
\label{eq:landscape}
\ee
The adimensional parameter $\a(\S)$ is related to the energy gap of the
ground state $\C^*$ of sequence $\S$, as defined by Shakhnovich and
coworkers \cite{gap}. However, it characterizes more precisely the smoothness
of the energy landscape.
For the initial sequence we find $\a(\S_0)=0.23$. We did not measure $\a(\S)$
for all selected sequences, but it appears from few examples that its value
does not decrease during the evolution of the protein.
It is not surprising that selected sequences exhibit a correlated energy
landscape, since the condition of thermodynamic stability that we impose
rules out sequences with misfolded structures of low energy. Moreover,
our selected sequences are fast folders, and one should expect that such
sequences have a correlated energy landscape, since the relation between 
thermodynamic stability, fast folding
and smoothness of the energy landscape has long been discussed \cite{BW,SGinv}.
Furthermore, it has been found that models which cannot give
rise to good folding sequences present weak correlations between $q$ and $E$
\cite{michele,Helge}.

Consistently, it appears from our data that the folding time is correlated to
the stability (it has a negative correlation with the $Z$ score and positive
with the native overlap), but we are not able to quantify this effect,
since our measure of the folding time, based only on two simulations,
is too imprecise.

It is also interesting that for several sequences the energy $E(\C^*,\S)$
is lower than for the starting sequence $\S^*$, although this
has been obtained by minimizing the energy $E(\C^*,\S)$ in sequence space.
The reason for this is that the minimization method requires that the
composition of the sequence is kept fixed, 
while we do not impose this
condition. However, the $Z$ score reaches its lowest
value $Z=-1.22$ for the starting sequence $S^*$.

\begin{figure}
\centerline{\psfig{figure=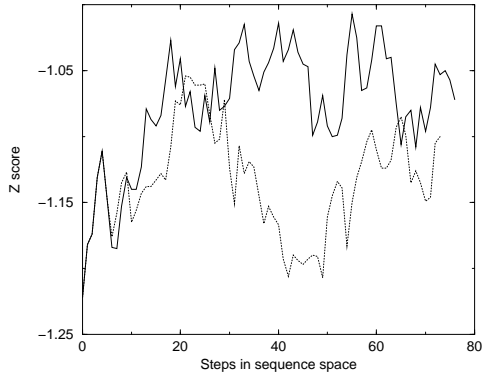,width=7.0cm,angle=0}}
\caption{$Z$ score of sequences in the neutral network as a function of the
number of steps along the network, starting from $\S^*$, for
two evolutionary trajectories.}
\label{fig:landscape}
\end{figure}

We observe that the $Z$ score of the target state, $Z(\C^*,\S)$, defines
a complex landscape in sequence space, with valleys separated by barriers.
This result is illustrated in Fig.\ref{fig:landscape}, where we show
the $Z$ score of sequences in the neutral network as a function of the
number of steps $l$ along the network, starting from $\S^*$.

The roughness of the energy landscape is consistent 
with the results reported in a recent preprint 
by Tiana {\it et al.} \cite{Tiana2}, who sampled
the energy landscape in sequence space for a fixed structure using Monte
Carlo simulations. The authors of \cite{Tiana2} observed a hierarchy of
clusters and superclusters of low energy sequences, with superclusters
characterized by few fixed amino acids in key positions and not connected
by neutral paths.
This conclusion might at first seem at odd with the fact that we found
connected neutral paths extended in sequence space,
but there should be no contradiction between the present work and
the results of ref. \cite{Tiana2}, since they deal with different questions.
We studied a single neutral network asking whether it is possible to find
in it pairs of sequences at any distance,
while Tiana {\it et al.} \cite{Tiana2}
ask whether it is possible to find non homologous proteins which are in
disconnected neutral networks and still fold to the same structure.
It is possible that both answers are positive and that several extended
but disconnected neutral networks exist for a given protein structure.
Such a picture, if correct, implies that non homologous proteins sharing
the same fold may have been originated either through convergent
evolution,  possibly on disconnected neutral networks, or through
divergent evolution from a common anchestor on a single neutral network, but
it is very difficult or even impossible to decide between these two
possibilities on the basis of the sequence alone.

On the other hand, we should note that the definition of neutral path 
in our work is different than the one used in \cite{Tiana2}. In fact,
while in \cite{Tiana2} is assumed that a path in sequence space is neutral if 
all sequences
belonging to it have $Z$ score lower than a predetermined threshold, we
accept only sequences for which we can show through Monte Carlo simulations
that the target structure is the ground state, it is thermodynamically
stable and easy to reach kinetically. The criterion adopted in \cite{Tiana2}
has the advantage of being computationally very efficient and it correlates
well with our criterion. In several cases, however, the two criteria give
different answers, as it is shown in Fig. \ref{fig:threshold}, and it is
possible that networks which are disconnected according to the $Z$ score
criterion are found to be connected according to our criterion.
Further study is necessary in order to assess the relevance of
such possibility.

It is rather interesting that the complex energy landscape in sequence
space offers an explanation for the large variations of the neutral
mutation rate, $x(\S)$. Since a more stable sequence can tolerate larger
rearrangements without changing its ground state, one can expect that
the fraction of neutral mutations from sequence $\S$ increases with
the stability of the sequence. This expectation is in agreement with the
results of ref. \cite{Tiana2}, where the stability is measured by the
$Z$ score $Z(\C^*,\S)$ and neutrality of a mutation is recognized with a
criterion based on the $Z$ score of the mutated sequence.

\begin{figure}
\centerline{\psfig{figure=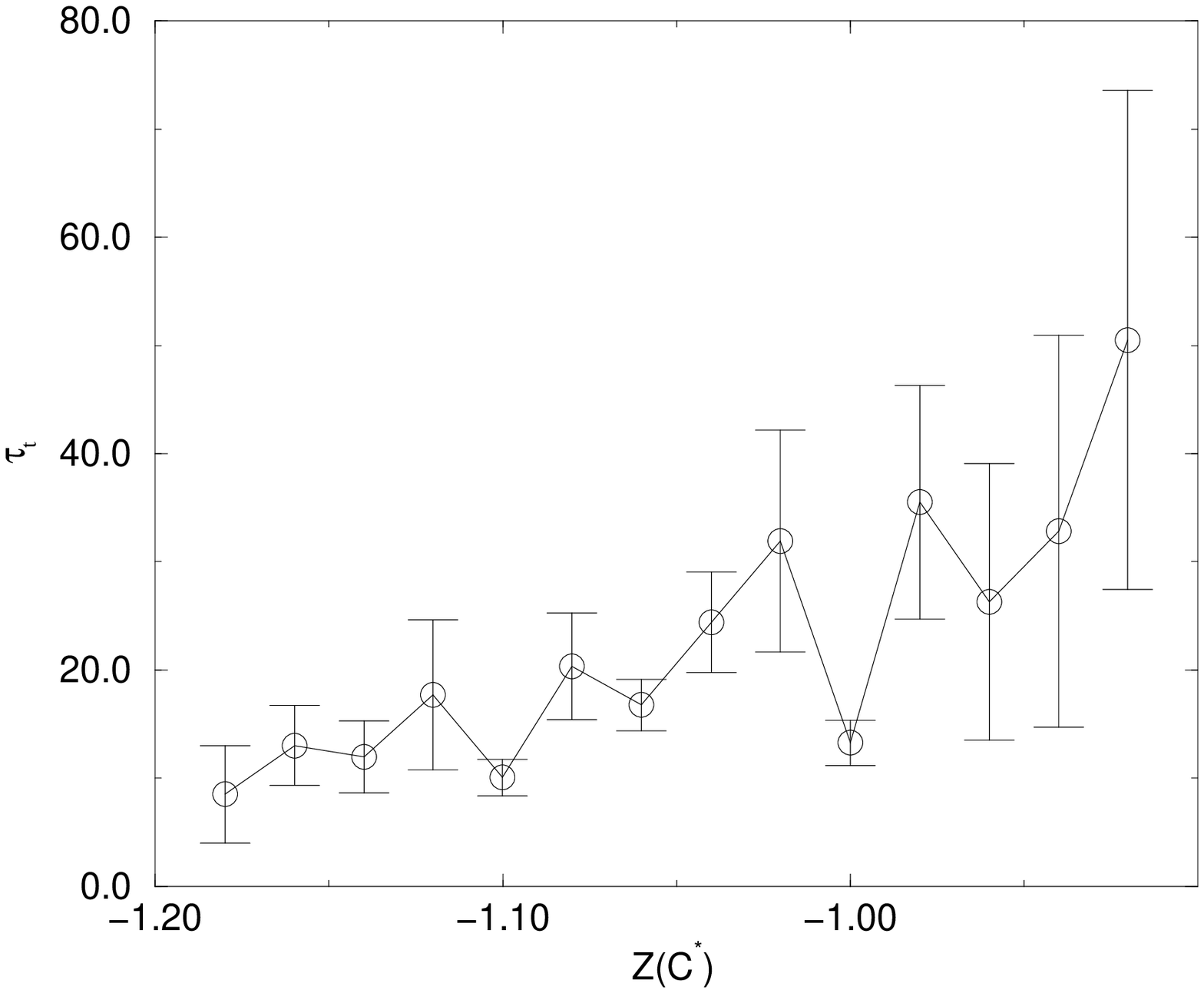,width=7.0cm,angle=0}}

\centerline{\psfig{figure=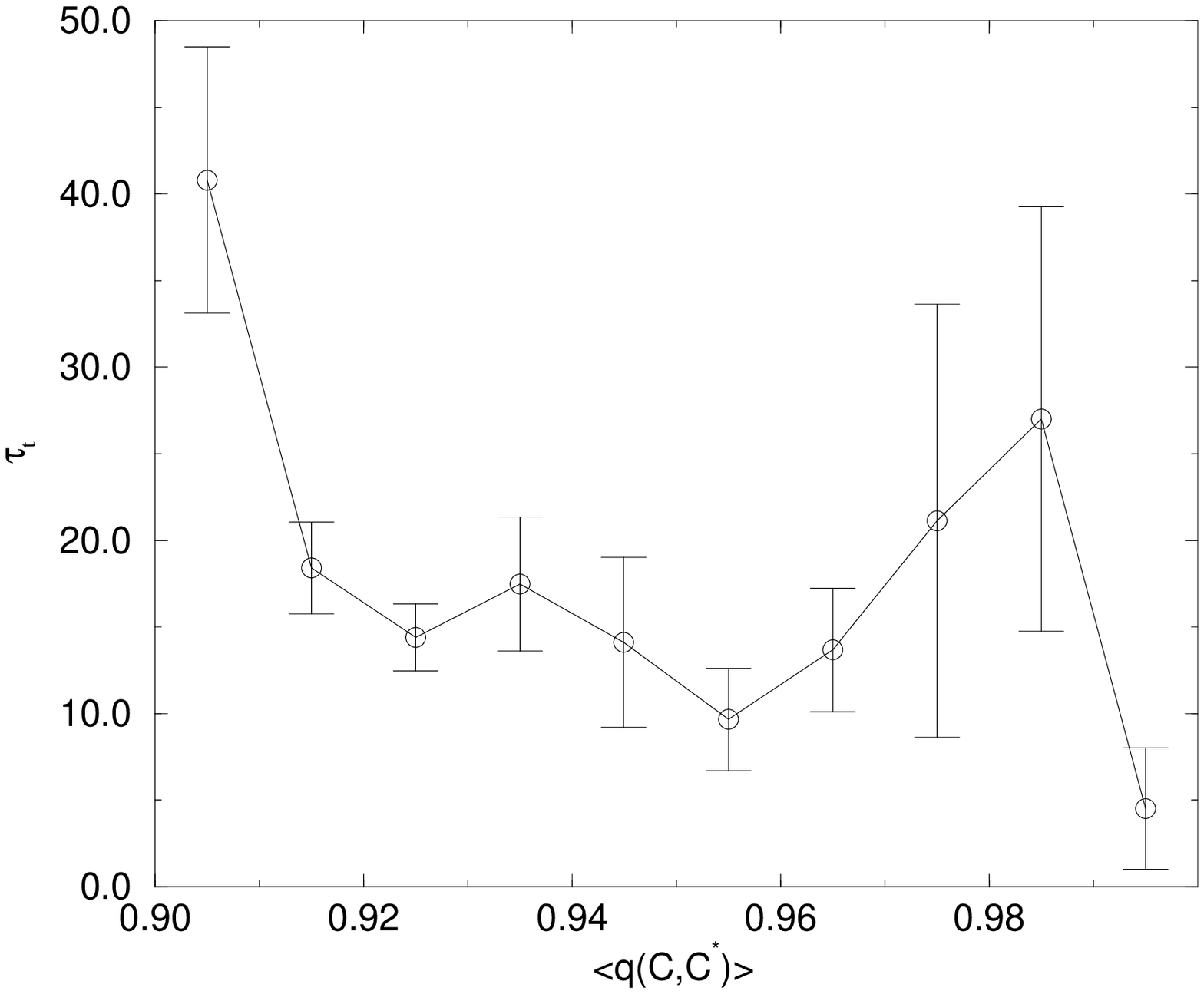,width=7.0cm,angle=0}}
\caption{Average value of $\tau_t$ for sequences of the neutral network
as a function of the $Z$ score (upper panel) and of the average native
overlap (lower panel).}
\label{fig:Ztau}
\end{figure}

We study the correlation between the stability, measured either by
the native overlap or by the $Z$ score, and the fraction of neutral
neighbors $x(\S)$. Since we did not measure $x(\S)$, we have to rely
on the trapping time $\t_t$ spent by a trajectory on sequence $\S$.
This variable is related to $x(\S)$ through the geometric
distribution $P_x(\t)=x(1-x)^{\t-1}$ of average value $1/x(\S)$
(\ref{eq:Ptau}). We thus estimate the correlation
coefficient between the $Z$ score and $1/x(\S)$, using the relations
$\l[Z(\S)/x(\S)\r]\approx \l[Z(\S)\tau_t(\S)\r]$,
$\l[1/x(\S)\r]\approx \l[\tau_t(\S)\r]$,
$\l[1/x(\S)^2\r]\approx 1/2\l[\tau^2_t(\S)+\tau_t(\S)\r]$,
where the square brackets indicate average on the neutral network.
This treatment neglects the fact that our criterion is subject to some
arbitrariness,
since we can not measure with high precision the native
overlap $\la q\ra$ and the typical folding time upon which our criterion
is founded. Thus, the correlation coefficient estimated in this way
is underestimated. We find a correlation coefficient $r=0.20$
between the $Z$ score and $1/x$ and $r=-0.21$ between the native overlap
and $1/x$. Although the estimate is not accurate, this
study confirms the existence of correlations between the stability of the
native state and the neutral mutation rate.
We show the correlation between $Z$ and $\t_t$ in Fig.\ref{fig:Ztau}.

The previous observation can be the basis for a quantitative explanation
of the large fluctuations of the neutral mutation rate in this model and
possibly in real proteins. It would be very interesting to investigate 
to which extent such fluctuations are responsible for the observed patterns of
molecular evolution, which appear to be much more irregular than 
predicted on the basis of the simple ``homogeneous'' neutral theory.

\subsection{Unstable and slow folding sequences}

To these two classes belong all sequences that were rejected although
no structure of energy lower than $\C^*$ was found.
We do not know for which fraction of these sequences the ground state
coincides with $\C^*$ and for which the ground state is changed.
For unstable sequences $\C^*$ is the lowest energy structure found and it is
reached in at least one of the two independent MC runs, but the stability
condition is not fulfilled. The rejection was made already in the first
MC run if we found $\la q\ra > 0.75$. Thus we can not exclude that in the
second run also the fast folding condition would fail. For slow folding
sequences $\C^*$ was not reached either in the first or in the second
MC run. The folding time for unstable sequences appears to be correlated
to the native overlap $\la q\ra$, even if our data do not allow
quantitative estimations.

\subsection{Structurally mutated sequences}

For sequences in this class we found putative ground states with energy
lower than that of the target structure. A fraction $f=0.306$ of the
examined sequences belongs to this class.
 
\begin{figure}
\centerline{\psfig{figure=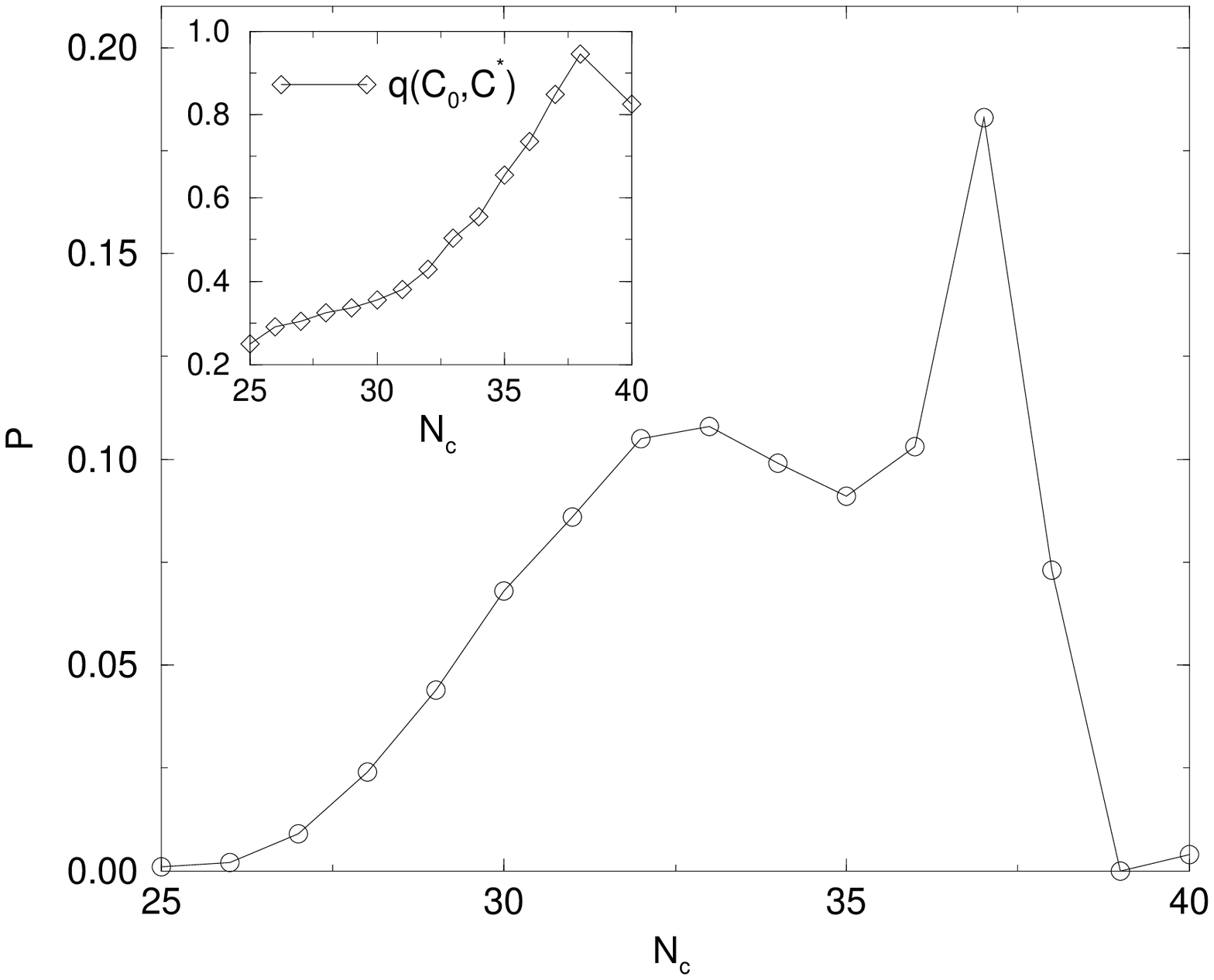,width=7.0cm,angle=0}}
\caption{Distribution of the number of contacts in the ground
  state, $N_c$, for all of the generated sequences. The peaks
  at high $N_c$ are related to target-like structures. Insert: correlation
between the number of contacts and the overlap $q_0$ between target state and
ground state.}
\label{fig:Nc}
\end{figure}

We first analyze the number $N_c^0$ of contacts in the mutated ground state. 
The distribution $P(N_c^0)$ is reminiscent of a bimodal distribution
(see Fig. \ref{fig:Nc}). As previously mentioned, the
number of target contacts, $N_c^*=40$, is the largest possible for a
sequence length of $N=36$ residues.
The twin peaks at $N_c^0=37$ and at $N_c^0=40$ derive from target-like
ground states, while the constraints of the lattice geometry 
are probably responsible for the depression of
the values of $P(38)$ and $P(39)$.
The broad peak at $N_c^0$=34 is close to the number of contacts expected
for random sequences, although slightly higher.
As mentioned above, in the case of a random contact potential with the
same mean and variance as the one that we used and a Gaussian distribution,
the number of contacts in the ground states ranges typically
from 29 to 33 \cite{ugo}.

The number of contacts is higher than for random
sequences because the ``native'' contacts of $\C^*$ are advantageous even 
in the structurally mutated sequences.
This is confirmed by the fact that there is a strong correlation
correlation between the number of contacts $N_c^0$ and the overlap
$q_0$ between the new ground state and the target state $\C^*$
The correlation coefficient is $r=0.80$. See insert in Fig.\ref{fig:Nc}.
The two quantities are also correlated to the energy $E_0$ of the ground
state: The more native-like the mutated ground state is,
the more compact it is and the lower is its energy. The correlation
coefficients are: $r=-0.31$ between $q_0$ and $E_0$, $r=-0.32$ ($N_c^0$
and $E_0$), $r=-0.54$ ($q_0$ and $Z$ score) and $r=-0.47$ ($N_c^0$ and $Z$
score). The strongest correlation observed is that between $q_0$ and the
$Z$ score. The energy and the $Z$ score are weekly correlated ($r=0.36$).

\begin{figure}
\centerline{\psfig{figure=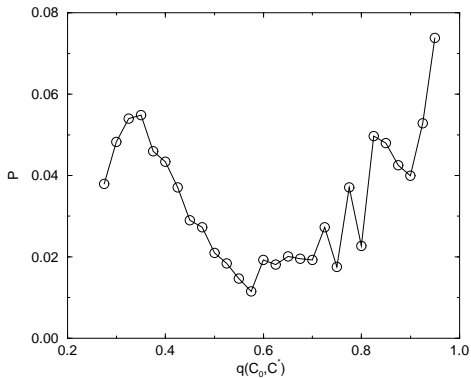,width=7.0cm,angle=0}}
\caption{Distribution of the overlap $q(\C^*,\C)$ between the target
  structure $\C^*$ and the ground-state of the sequences studied, $\C$.}
\label{fig:P_q}
\end{figure}

The distribution of $q_0$ is also bimodal (see Fig. \ref{fig:P_q}). 
The peak at high $q$ is due to native-like structures and the broader peak at
$q=0.3$ is close to (but still significantly higher than) the typical overlap 
between random structures, $q_{ran}=0.1$ \cite{ugo} for chains of
length $N=36$.

\begin{figure}
\centerline{\psfig{figure=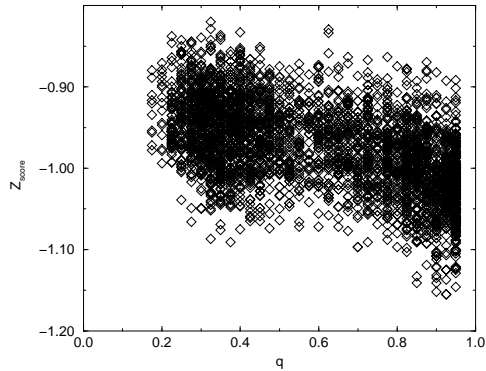,width=7.0cm,angle=0}}
\caption{Scattering plot showing the $Z$ score of the ground state as
a function of its overlap with the target state.}
\label{fig:mut}
\end{figure}

The overlap $q_0$ between the ground state and the target state is
negatively correlated to the $Z$ score. This is shown in Fig.\ref{fig:mut},
where all the structurally mutated sequences are represented in a
scatter plot in the plane $(q_0,Z)$. Only structures which are very
similar to the original native state have a $Z$ score in the range found
for selected sequences. This result suggests that mutated ground states
dissimilar from the original one are in most cases not stable enough to
represent a new acceptable fold of the model protein. In other words,
neutral networks of unrelated structures may lay far apart in sequence
space.
This feature of the model is consistent with the observation that
biological evolution conserves the native fold of proteins even when their
function changes substantially \cite{holm,qasba}.

\section{Discussion}
\label{sec:discussion}

We simulated evolution on neutral networks
for a protein model with twenty amino acid types, contact energy
functions and structures represented as self avoiding walks on the simple
cubic lattice. Stability of the ground state is measured as the thermodynamic
average of its overlap with alternative structures.

Our simulations show that neutral networks are extended in sequence
space: pairs of sequences on the neutral network are almost as different as
random sequences, even if they have exactly the same fold. This observation
is consistent with what is known about protein evolution.
In our 36mer chains, residues in all positions could be substituted, even if
the two positions in the core are much more difficult to change, consistently
with the results of Ref. \cite{ms99}.

The neutral network that we studied turned out to be rather irregular:
the fraction $x(\S)$ of neutral mutations of a sequence $\S$ has large
variations in sequence space. The value of $x(\S)$ is positively
correlated with the stability of the native state, evaluated either through
the average native overlap $\la q(\C,\C^*)\ra$ or through the $Z$ score.
This appears very reasonable: The more stable is a sequence to structure
match, the less probable is that a mutation destabilizes it.
This is also in agreement with the results of recent complementary
studies \cite{stability,Tiana3}. It thus seems that thermodynamic stability
implies also stability with respect to mutations.

The fluctuations of $x(\S)$ in sequence space are important for the
evolutionary dynamics. If $x(\S)$ is constant, Kimura's theory
of neutral evolution predicts that the overlap in sequence space relaxes
exponentially to the asymptotic value corresponding to random sequences, at a
rate independent of the size of the biological
population in which the evolution takes place. Moreover, the number of
substitutions in a protein sequence during an evolutionary trajectory
lasting $T$ generations is predicted to be a Poissonian variable of
mean value $\mu xT$. 
In this case, the fluctuations of the value of this variable
in different species evolving for a time $T$ should grow only as $\sqrt T$.
This result 
would sustain
the molecular clock hypothesis, according to which
the substitution process can be used as a clock to date speciation events.
However, the molecular clock hypothesis has been heatedly debated in the last
decade, and it has been shown that the number of substitutions
fluctuates much more than predicted in the ``homogeneous'' neutral theory
\cite{Gillespie}. This
deviation from the prediction of Kimura's theory has been interpreted as
an indication that in most cases protein evolution is not neutral but
adaptive. We suggest that the irregularity of protein evolution could 
be an intrinsic property of the energy landscape of neutral networks of
protein sequences. More stringent statistical tests should be designed
to distinguish this situation from adaptive evolution, that undoubtedly
occurs in many cases.

Selected sequences have a rather correlated energy landscape, which yields 
short folding times 
and high thermodynamic stability. The native
overlap that we use as a measure of thermodynamic stability correlates
well with the $Z$ score but it gives more information on the absence
of states of low energy unrelated to the native state and favors a more
correlated energy landscape.

Sequences whose ground state coincides
with the native state may be discarded either
due to the lack of thermodynamic
stability or because they fold too slowly. Both classes of sequences have
similar properties, since stability and folding time are related
quantities.

For about 30\% of the attempted mutations the resulting
sequence has a ground state different than the original one. The overlap
$q_0$ of this new ground state with the target state has a bimodal
distribution, but only structures very similar to the target one appear to
fulfill our criterion of thermodynamic stability. This is not surprising,
since the target structure must conserve a low energy in the mutated
sequence, so that it is able to destabilize the new ground state.
Therefore, it seems that contacts between neutral networks of unrelated
structures are very rare, if a stability condition is required. A
similar conclusion has been suggested in a numerical study of the two
dimensional HP model \cite{BB}.

An implication of this result is that it is difficult to switch from
a structure to a different one through point mutations corresponding
to thermodynamically stable proteins. This could explain why evolution
changes so rarely the fold of a protein, while it is possible to
engineer protein sequences with as much as 60\% similarity with a natural
protein and completely different fold.

\vspace{.2cm}
We studied a model which has the advantage of being
reliably computable, but at the price of sacrificing
possibly important ingredients.
We discuss here the ones that we judge the most serious:

\begin{enumerate}

\item 
We use a simple lattice model. This choice was 
made due to the necessity of identifying the native state of each generated 
sequence, and this is feasible only for lattice models.
Lattice models, although often criticized \cite{honig}, have been
recognized to capture some of the most relevant thermodynamic features of the
folding process \cite{s96}, such as the existence of a unique
ground state and the cooperativity of the transition. 
However, they do not capture essential features of
real proteins as for instance the existence of secondary structures.

\item 
We simulated the evolution of only one target structure. It
would be interesting to see how our results change by changing the
structure, and which properties of the structure (for instance
compactness, locality of interactions and so on) are important to determine
the neutral mutation rate. However, it was argued that the small
number of folds occurring in natural proteins (at most some thousands) could
be the ones corresponding to the largest number of sequences in
sequence space \cite{fold}, so that structures characterized by a
large neutral set, even if they are not typical, could be the most
interesting ones from the biological point of view.

\item 
The length of the sequences examined is short, so that there are
only two core residues. Considering more core residues could impose
more constraints on the evolution and reduce the rate of neutral
evolution. It would thus be interesting to make the same study for
longer sequences.

\item 
We did not represent biological activity in the present protein model.
This might be obtained by imposing more constraints on the
residues taking part to the active site.

\item 
In our model of evolution we assume that the environment remains
fairly constant, so that the native structure favored by natural
selection does not change throughout the evolution. This hypothesis
is not unreasonable if the protein examined is an enzyme
performing some chemical activity, since the cells possess a high
homeostasis, {\it i.e.} they can maintain a stable chemical-physical
internal environment despite large perturbations in the external
environment. However, it is quite likely that some large ecological
and climatic changes have been responsible for molecular
substitutions for which the neutral theory, and our model in
particular, do not apply \cite{Gillespie}.

\item 
We consider only point mutations, and not insertions and
deletions, which also play an important role in evolution.

\end{enumerate}

In our opinion, these limitations should not modify
the qualitative picture. The existence of neutral networks,
the variability of neutral mutation rates and the difficulty to reach
through point mutations very different structures 
corresponding to stable proteins 
are features of our model that appear to be reflected also in
the evolution of real proteins.

\section*{Acknowledgments}

We acknowledge interesting discussions with 
Peter Grassberger, Helge Frauenkron, Erwin Gerstner,
Walter Nadler, Peter Schuster, Anna Tramontano,
Tim Gibson, Erich Bornberg-Bauer, Guido Tiana, Ricardo Broglia and 
Eugene Shakhnovich.
This work was conceived during
the workshop on Protein Folding organized at the ISI Foundation, Torino,
Italy, February 9-13 1998.
Part of the work was made during the Euroconference on "Protein
Folding and Structure Prediction" organized at the ISI Foundation,
Torino, Italy, June 8-19, 1998.
Computations were carried out at the HLRZ,
Forschungszentrum J\"ulich.


\begin{thebibliography}{99}
\bibitem{random}
The number of amino acid matches obtained by pairing
two random sequences of the same length is given by
the binomial distribution
with $p=1/20$ if one assumes that the 20 amino acids have the
same probability to occur.
For sequences of length $N$ there will be on average
$pN$ identical amino acids, with a variance $Np(1-p)$.
For random sequences, 95\% of pairwise comparisons
yield a sequence identity between 1\% and 9\%.
\end{thebibliography}
\end{document}